\documentclass[12pt,final]{iopart}

\usepackage{geometry}
\usepackage[parfill]{parskip}    
\usepackage{graphicx}
\usepackage[ansinew]{inputenc}
\usepackage{palatino, url, multicol}
\usepackage[usenames,dvipsnames]{color}
\usepackage{iopams}  
\usepackage{subfigure}
\DeclareGraphicsExtensions{eps}
\usepackage[center]{caption}

\newcommand{\citeb}[1]{\textcolor{blue}{\cite{#1}}}
\newcommand{\red}[1]{\textcolor{red}{#1}}

\begin{document}

\title[]{On the performance of Usain Bolt in the 100 metre sprint}

\author{JJ Hern\'andez G\'omez, V Marquina and RW G\'omez}

\address{Facultad de Ciencias, Universidad Nacional Aut\'onoma de M\'exico.\\ Circuito Exterior C.U. M\'exico D.F., 04510, M\'exico}
\eads{\mailto{jorge\_hdz@ciencias.unam.mx}, \mailto{marquina@unam.mx}, \mailto{rgomez@unam.mx}}

\begin{abstract}
Many university texts on Mechanics deal with the problem of the effect of the air drag force, using as example the slowing down of a parachute. Hardly no one discuss what happens when the drag force is proportional to both $u$ and $u^2$. In this paper we deal with a real problem to illustrate the effect of both terms in the speed of a runner: a theoretical model of the performance of the 100 m world record sprint of Usain Bolt during the 2009 World Championships at Berlin is developed, assuming a drag force proportional to $u$ and to $u^2$. The resulting equation of motion is solved and fitted to the experimental data obtained from the International Amateur of Athletics Federations that recorded Bolt’s position with a LAVEG (laser velocity guard) device. It is worth to note that our model works only for short sprints. \end{abstract}

\vspace{2pc}

\submitto{\EJP}
\maketitle

\section{Introduction}

In June 21, 1960, at Zürich, Switzerland, the German Armin Harry astounded the sports world achieving what was considered the physiological and psychological barrier for the 100 m dash: 10 s flat. It was until June 20, 1968, at Sacramento, California, that Jim Hines ran 100 m in 9.9 s, breaking this barrier. In following years many sprinters had run this distance faster than 10 s, but 31 years were needed to lower Harry’s record by 0.14 s (Carl Lewis, August 25, 1991, at Tokyo, Japan). The current world record of 9.58 s was established by Usain Bolt (who also held the 200 m world record of 19.19 s up to 2012) in the 12th IAAF World Championships in Athletics at Berlin, Germany (2009). 

The performance of Usain Bolt in the 100 meter sprints is of physical interest since he can achieve, until now, accelerations and speeds that no other runner can. Through time, several mathematical models to fit the position, the velocity, or both, of a sprinter have been proposed \citeb{Keller,Alexandrov,Tibshirani,Wagner,Heck,Helene}. Recently, Helene \etal \citeb{Helene} fitted Bolt’s performance during both the summer Olympics in 2008 at Beijing and the world championships in 2009 at Berlin, using a simple exponential model for the time dependence of the speed of the runner. 

\section{Theoretical model} \label{S2}


The important forces acting during the race are the horizontal force that Bolt exerts and a drag force that depends upon the horizontal velocity (speed). Other factors, such as humidity, altitude above sea level (36 m), oxygen intake and that he turns his head to watch other runners, affecting the mechanics of his motion, are not taken into account. Based on the fact that Bolt’s 200 m time is almost twice the one for 100 m, our main assumption is that in the 100 m dash, he is able to develop a constant horizontal force $F_0$ during the whole sprint. The drag force, $D(u)$, is a function of Bolt’s horizontal speed respect to ground $u(t)$, with or without wind. This force causes a reduction of his acceleration so his speed tends to a constant value (terminal speed). Thus, the equation of motion is
\begin{equation}
m\dot{u}=F_0-D(u).  \label{eq1}
\end{equation}

This equation can be readily casted as a quadrature,
\begin{equation}
t-t_0=m\int_{u_0}^u \frac{du'}{F_0-D(u')} . \label{eq2}
\end{equation}

The integral above does not have an analytical solution for a general drag function; however the drag force can be expanded in Taylor series,
\begin{equation} 
D\left( u \right)\simeq D\left( 0 \right)+{{\left. \frac{dD\left( u \right)}{du} \right|}_{0}}u+\frac{1}{2}{{\left. \frac{{{d}^{2}}D\left( u \right)}{d{{u}^{2}}~} \right|}_{0}}{{u}^{2}} +O\left(u^3\right) . \label{eq3}
\end{equation}

The constant term of the expansion is zero, because the runner experiments no drag when at rest. The second and third terms must be retained. While the term proportional to the speed represents the basic effects of resistance, the term proportional to the square of the speed takes into account hydrodynamic drag, obviously present due to the highly non-uniform geometry of the runner. In general, for relatively small speeds, it suffices to take only the first three terms of the expansion. 

Renaming the $u$ and $u^2$ coefficients as $\gamma$ and $\sigma$ respectively, the equation of motion \eref{eq1} takes the form
\begin{equation}
m\dot{u}={{F}_{0}}-\gamma u-\sigma {{u}^{2}}, \label{eq4}
\end{equation}
whose solution follows straightforward from \Eref{eq2},
\begin{equation}
u\left( t \right)=\frac{AB\left( 1-{{e}^{-kt}} \right)}{A+B{{e}^{-kt}}} , \label{eq5}
\end{equation}
where the coefficients are related by $\sigma =km/(A+B)$, $F_0=kmAB/(2A+2B)$ and $\gamma=km(A-B)/(A+B)$.

The position can be obtained by integrating \Eref{eq5},
\begin{equation}
x(t)=\frac{A}{k}\ln\left( \frac{A+Be^{-kt}}{A+B} \right)+\frac{B}{k}\ln\left( \frac{Ae^{kt}+B}{A+B} \right) , \label{eq6}
\end{equation}
while the acceleration can also be calculated by deriving \Eref{eq5},
\begin{equation}
a(t)=ABk(A+B)\frac{e^{-kt}}{\left( A+Be^{-kt} \right)^{2}} . \label{eq7}
\end{equation}

\section{Experimental data fitting}

\begin{figure}[h!]
\centering \subfigure[ ]    {
     \includegraphics[width=0.48\textwidth]{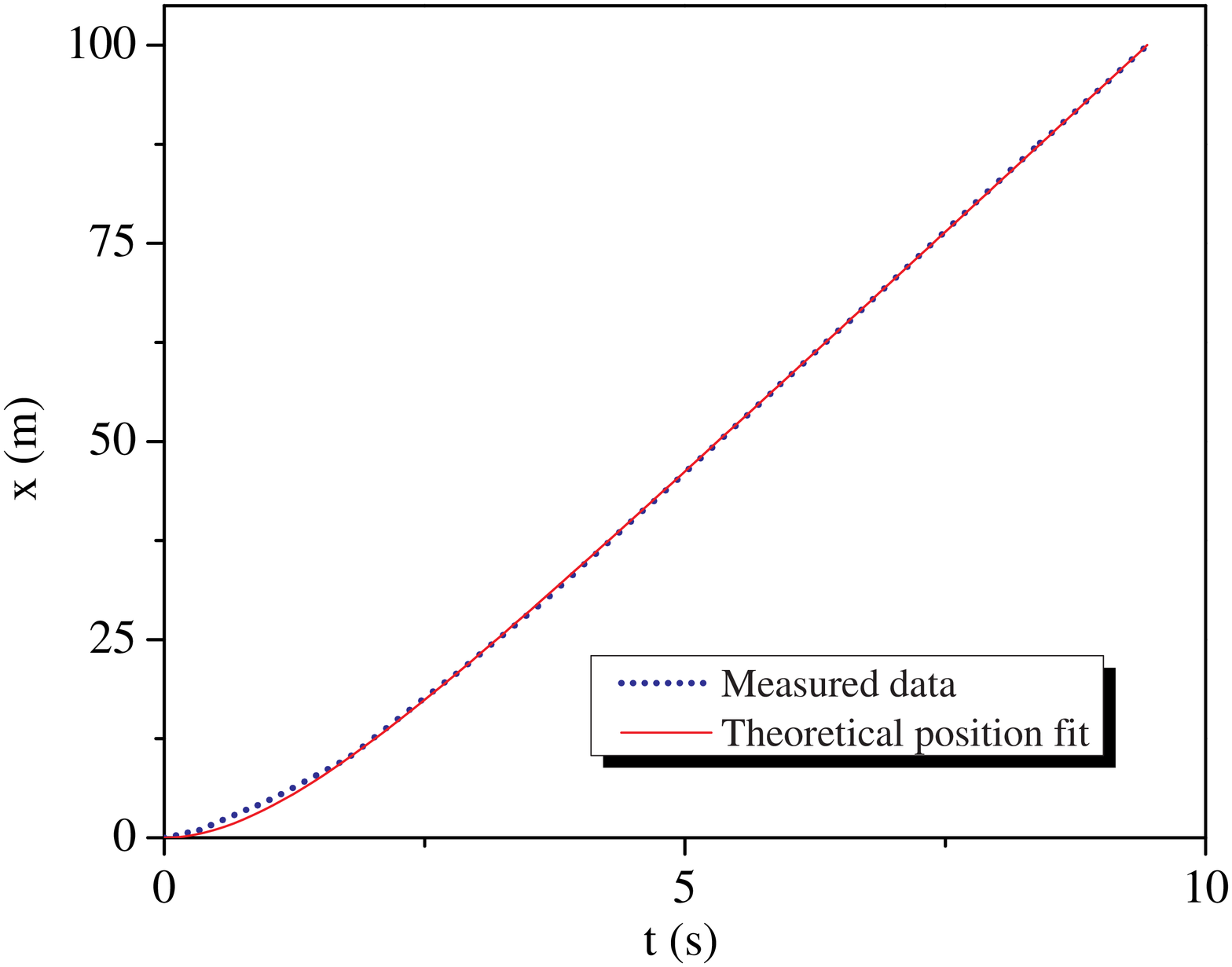} 
	\label{f1a} 
    }    \subfigure[ ]    {
     \includegraphics[width=0.48\textwidth]{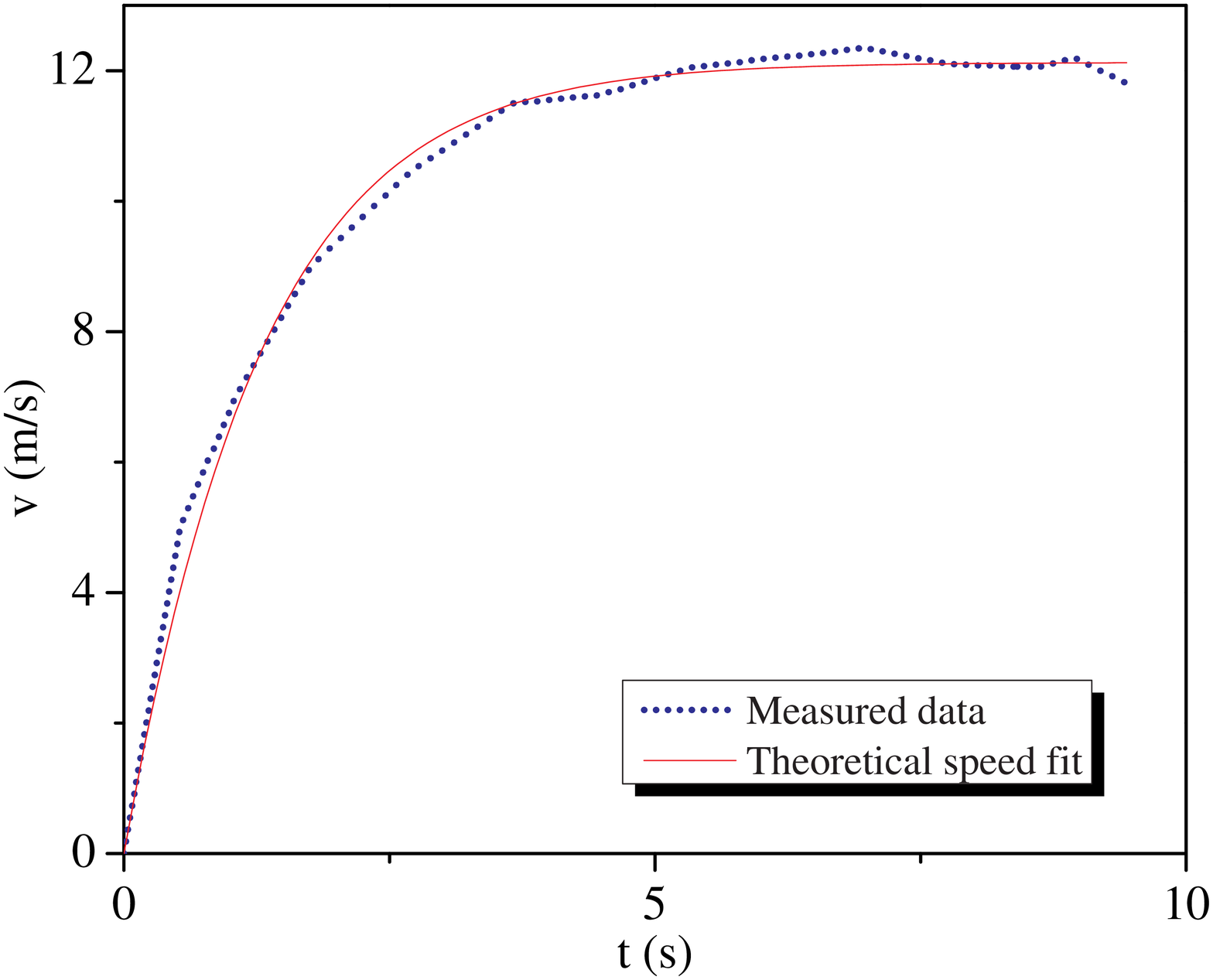} 
	\label{f1b} 
    }\caption{Position (a) and speed (b) of Bolt in the 100 m sprint at the 12th IAAF WCA. The dotted (blue) line corresponds to the experimental data while the solid (red) one corresponds to the theoretical fitting.}
    \label{f1}
\end{figure}

The experimental data we used were from the 12th IAAF World Championships in Athletics (WCA), which were obtained from \citeb{Video}, and consist of Bolt’s position and speed every $1/10$ s. To corroborate the accuracy of the data obtained from \citeb{Video}, we reproduced with them the velocity vs position plot given in \citeb{IAAF}, which was obtained by the IAAF, by means of a LAVEG (laser velocity guard) device. The parameters $A$, $B$ and $k$ were fitted by a least-squares analysis, with the Origin 8.1 © software, in both position and speed data sets, considering a reaction time of  0.142 s \citeb{Helene}. In \fref{f1a} and \fref{f1b} we show such fittings, together with the experimental data.

The parameter values for both fittings are shown in \tref{t1a}. We do not report errors, because the standard error of the fitting on each parameter lies between the second and the third significant digit, which is finer than the measurement error in the data.

\begin{table}[h!]
\centering
\caption{Fitted values of the parameters $A$, $B$ and $k$.}
\begin{tabular}{ccc}
\br
   \textbf{Parameter}   &  \textbf{Position fitting}  & \textbf{Velocity fitting}\\
\mr
$A$ (m/s)   &  110.0  &  110.0  \\
$B$ (m/s)   & 12.2   &  12.1	  \\
$k$ (1/s)    &  0.9  &   0.8 \\
\br
\end{tabular}
\label{t1a}
\end{table}

\begin{table}[h!]
\centering
\caption{Values of the physical parameters $F_0$, $\gamma$ and $\sigma$.}
\begin{tabular}{cc}
\br
   \textbf{Constant}   &    \textbf{Value}   \\
\mr
$F_0$ (N)   &  815.8  \\
$\gamma$ (kg/s)  &  59.7    \\
$\sigma$ (kg/m)  &  0.6  \\
\br
\end{tabular}
\label{t1b}
\end{table}

The accuracy of the position and velocity fittings is $R_p^2=0.999$ and $R_v^2=0.993$ respectively, so we decided to use from now on the results of the parameters$A$, $B$ and $k$ from the position fitting. The computed values of the magnitude of the constant force $F_0$, and the drag coefficients, $\gamma$ and $\sigma$, are shown in \tref{t1b}, taking Bolt’s mass as 86 kg \citeb{Charles}.

\begin{figure}[h!]
\centering     \includegraphics[width=0.50\textwidth]{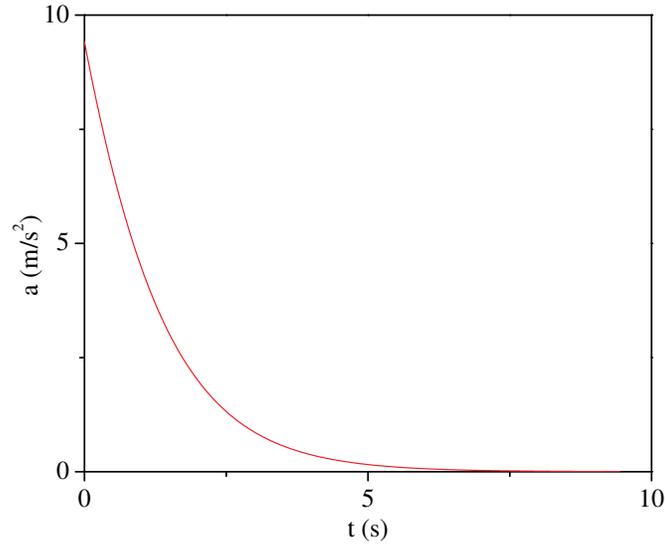} 
\caption{Theoretical acceleration of Bolt in the 100 m sprint at the 12th IAAF WCA.}
\label{f2}
\end{figure}

We also show in \fref{f2} the plot of the magnitude of the acceleration we obtained; no fitting was made because there are no experimental data available.

\section{Results}

As any mechanical system subject to drag, the runner experiments a terminal velocity $u_T$ which is formally obtained when $\dot{u}=0$ in the equation of motion \eref{eq1}; that is, by solving the equation
\begin{equation}
F_0=D(u_T) \label{eq8}
\end{equation}
for $u_T$. Nevertheless, the solution of the equation for the terminal velocity can also be found when $t\to \infty$ in \Eref{eq5}, and it turns out to be $u_T=B$. Therefore, under this model, the runner acquires a terminal speed of $u_T=12.2$ m/s, which is physically feasible (see \fref{f1b}). According to the data obtained from \citeb{Video}, the average speed in the second half of the sprint, which is surprisingly equal to 99\% of the maximum speed recorded \citeb{Video}, is $12.15$ m/s. Moreover, the initial acceleration of Bolt is $a(0)=9.5$ m/s, which is of the order of the acceleration of gravity, $g$; this value of the initial acceleration is fully reasonable, considering that the acceleration a man must exert in order to be able to jump half of his own height, should be just slightly greater than $g$. Furthermore, the value of the constant force in \tref{t1b}, $F_0=815.8$ N, is totally consistent with the fact that one expects that the maximum constant (horizontal) force he could exert should be of the order of his weight, i.e. $w=842.8$ N.

Now, $\sigma=0.5 \rho C_d A$ represents the hydrodynamic drag, where $\rho$ is the density of air, $C_d$ the drag coefficient of the runner and $A$ his cross section area. The density of air at the time of the spring can be approximated as follows. Berlin has a mean altitude of 34 m over the sea level, and an average mean temperature for the month of August\footnote{The sprint took place on August 16, 2009.} \citeb{WMO} of 18.8 °C. Bearing in mind that the race took place at night, we consider an average temperature between the average mean temperature and the mean daily minimum temperature for August at Berlin, which is 14.3°C. Thus, the density of air is $\rho=1.215$ Kg/m$^3$ and the drag coefficient of Bolt is $C_d=2\sigma$/$\rho A=1.2$, where the cross section area of Bolt\footnote{To calculate such a cross section area, we used a similar procedure to the one used in \citeb{Charles}, where instead of a circle we estimated the area of the head with an ellipse. We averaged several scaled measures from Bolt pictures taken from \citeb{Bolt}} was estimated as $A=0.8$ m$^2$. This value of $C_d$ lays in the typical range for human beings reported in the literature (between 1.0 and 1.3) \citeb{Shanebrook,Zatsiorsky,Brownlie}.

The instantaneous power that Bolt develops, considering the drag effect is simply
\begin{equation}
P(t)=Fu=m\dot{u}u=mABk(A+B)\frac{(1-e^{-kt})e^{-kt}}{\left( A+B e^{-kt} \right)^3} . \label{eq9}
\end{equation}

\begin{figure}[h!]
\centering     \includegraphics[width=0.50\textwidth]{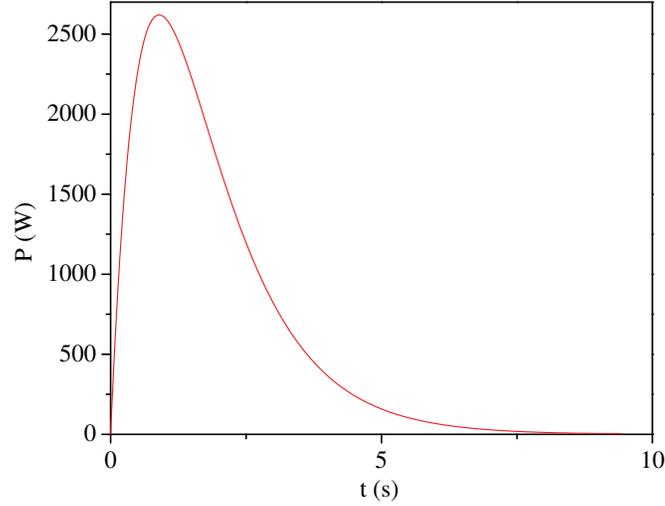} 
\caption{Theoretical power of Bolt in the 100 m sprint at the 12th IAAF WCA.}
\label{f3}
\end{figure}

In \fref{f3} we plot the power of the sprint for Bolt and the drag. It is  remarkable that the maximum power of ${{P}_{max}}=2619.5$ W (3.5 HP) is reached at a time of 
${{t}_{Pmax}}=0.89$ s, when the speed $u({{t}_{{{P}_{\max }}}})=6.24$ m/s was only about half of the maximum speed. The fact that the maximum instantaneous power arises in such a short time indicates the prompt influence of the drag terms in the dynamics of the runner. 

The effective work (considering the effect of the drag force) is then
\begin{equation}
W_\mathrm{Eff}=\int_0^\tau P(t)dt=\int_0^\tau\frac{1}{2}mdu^2=\frac{1}{2}mu^2 (\tau), \label{eq10}
\end{equation}
where $\tau$ is the running time (the official time of the sprint minus the reaction time of the runner). The effective work is the area under the curve of \fref{f3}, and it is $W_\mathrm{Eff}=6.36$ kJ. On the other hand, as Bolt is assumed to develop an essentially constant force, his mechanical work is simply $W_\mathrm{B}=F_0d=81.58$ kJ, where $d$ is the length of the sprint (100 m). This means that from the total energy that Bolt develops, only 7.79\% is used to achieve the motion, while  92.21\% is absorbed by the drag; that is, 75.22 kJ are dissipated by the drag, which is an incredible amount of lost energy.

\section{Discussion}

\begin{figure}[h!]
\centering     \includegraphics[trim=0cm 8cm 0cm 0cm,clip=true,width=0.70\textwidth]{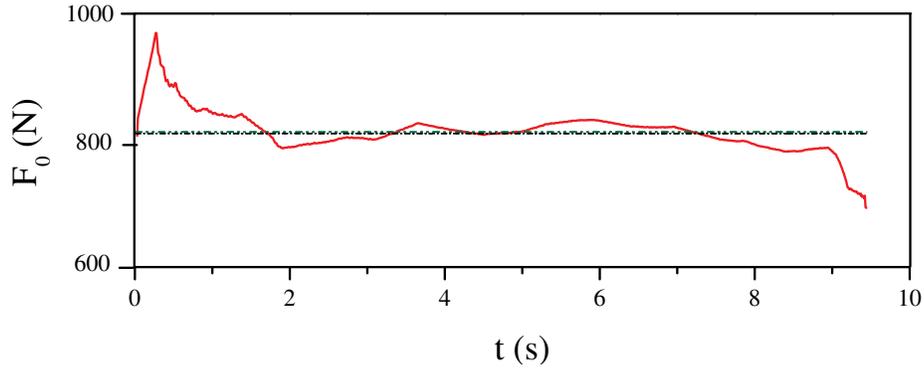} 
\caption{Force exerted by the runner during the race. The red line is calculated with the experimental data, the dash-dot-dot (green) line is the average force of 818.3 N, while the short-dash-dot (black) line is the value of the force $F_0$ obtained from the adjustment (815.7 N).}
\label{f4}
\end{figure}

As mentioned in section \ref{S2}, a central assumption in our model is that a 100 m sprinter (not only Bolt) is able to develop a constant force during the race (except in the initial few tenths of a second where he pushes himself against the starting block). In order to delimit how good is this assumption, we use the experimental values of $u$, the calculated acceleration, and the fitted values of the constants $\gamma$ and $\sigma$, to compute $F_0$. The result is shown in figure \ref{f4}. It is interesting to note that the average value of the force obtained from this figure is 818.3 N, which is very close to the value obtained from the fitting of the data\red{,} 815.7 N. The high value of the force in the first tenths of the race is due do the acceleration he obtains when he pushes himself from the starting block.

At a first glance, observing the values of the drag coefficients in \tref{t1b}, one is impelled to argue that, because $\sigma \ll \gamma $, the hydrodynamic drag could have been neglected. However, one can calculate the drag terms in the equation of motion at the terminal speed 
$u_T$, attaining $\gamma u_T=725.59$ N and $\sigma {u_{_{T}}}^2=90.18$ N. Thus, from the total drag $\gamma u_T+\sigma u^2_T$, 11.05\% corresponds to turbulent drag, which turns to be an important contribution.

If we would like to make predictions considering different wind corrections, this can be done as follows. Once a runner acquires the wind speed (which is almost instantly), the second term in the right side ($\gamma u$) of equation \eref{eq1} behaves as if the sprinter would be running in still air, because  $\gamma$ is proportional to the air viscosity, which is independent of air pressure. However that is not the case for the third term in \eref{eq1}, ($\sigma u^2$), which arises from the collisions per unit time of the air molecules against the sprinter and it is proportional to the speed of the runner respect to ground. In a simple model, the hydrodynamical drag force is $D_H =\sigma(\rho)(v + v_w)^2$, where $v$ is the speed achieved by the runner without wind and $v_w$ is the speed of the wind. The value of $\sigma$ depends on the number of molecules that impact on the runner per unit time and should be different in still air conditions. Then, the equation of motion \eref{eq1} can be rewritten as
\begin{equation} \label{eq11}
m\dot{u}=m\dot{v}={{F}_{0}}-\gamma v-\sigma {{(v+{{v}_{V}})}^{2}},
\end{equation}
and without wind as
\begin{equation} \label{eq12}
m\dot{v}={{F}_{0}}-\gamma v-{\sigma }'{{v}^{2}}.
\end{equation}
Subtracting \eref{eq11} and \eref{eq12}, we obtain
\begin{equation} \label{eq13}
\sigma ({{v}^{2}}+2v{{v}_{w}}+v_{w}^{2})=\sigma'v^2 ,
\end{equation}
so then
\begin{equation}  \label{eq14}
\sigma'=\sigma \left( 1+\frac{2{v}_{w}}{v}+\frac{v_w^2}{v^2} \right)  \sim \sigma \left( 1+\frac{2{v}_{w}}{v}\right) ,
\end{equation}
where the third term in the second expression has been neglected ($v_w<<v$). In order to estimate the value of $\sigma\rq{}$, we consider $v$ as the terminal speed of Bolt, $u_T$. With these conditions, $\sigma\rq{}$ = 0.69 with still air ($v_w = 0$ m/s) and $\sigma\rq{} = 0.49$ with a tailwind of $v_w = 2$ m/s. It should be clear that the present calculation is only a crude way to estimate the differences of running time with and without wind. The results, which are close to the values reported in literature\citeb{Mureika}, are summarized in \tref{t2}.
\begin{table}[h!]
\centering
\caption{Predictions of the running time for Bolt without tailwind, and with a tailwind of 2\nobreak m/s.}
\label{t2}
\begin{tabular}{cc}
\br
$\boldsymbol{v_w}$ \textbf{(m/s)}   &  \textbf{Estimated running time (s)}\\
\mr
 0 &  9.68   \\
 0.9 &  9.58   \\
 2 &  9.46  \\
\br
\end{tabular}
\end{table}

Although this is a simple way to calculate a correction due to wind, it turns to be a good proposal for it. A more realistic assumption would be to modify equation \eref{eq14} to be 
\begin{equation}  \label{eq15}
\sigma'=\sigma \left( 1+\frac{\alpha{v}_{w}}{u_T}\right) ,
\end{equation}
with the parameter $\alpha$ lying between 1 and 2. 

The results we obtained, altogether with the facts pointed out in this discussion, shows the appropriateness and quality of the model developed in this paper. We look forward for the next IAAF WCA, which will be held in Moscow, Russia, from August 10 to August 18, 2013, to test our model with the experimental data obtained from such sprints, as well as to wait expectantly if the fastest man on earth is able to beat his own world record once again.

\ack

This work was partially supported by PAPIIT-DGAPA-UNAM Project IN115612.

\end{document}